\documentclass[times, 5p, sort&compress, letterpaper]{elsarticle}

\usepackage{amsmath}
\usepackage{amssymb}

\usepackage[colorlinks]{hyperref}

\newcommand{\cov}{\textbf{Cov}}

\newcommand{\kv}{\textbf{k}}

\newcommand{\tpc}{(2 \pi)^3}

\newcommand{\taunl}{\tau_{\rm NL}}
\newcommand{\fnl}{f_{\rm NL}}

\newcommand{\Bl}{\left\langle}
\newcommand{\Br}{\right\rangle}

\newcommand{\lcdm}{\Lambda{\rm CDM}}
\newcommand{\slcdm}{{\rm Super-}\lcdm}

\newcommand{\kmsMpc}{{\,\rm km/s/Mpc}}
\newcommand{\lnB}{\ln B_{01}}

\journal{Physics of the Dark Universe}

\begin{document}

\begin{frontmatter}
\title{Super-CMB fluctuations and the Hubble tension} 
\author[1,2,3]{Saroj Adhikari} \ead{adh.saroj@gmail.com}

\author[1,2]{Dragan Huterer} \ead{huterer@umich.edu}    
\address[1]{Department of Physics, University of Michigan, 450 Church St, Ann Arbor, MI 48109-1040}
\address[2]{Leinweber Center for Theoretical Physics, University of Michigan,
  450 Church St, Ann Arbor, MI 48109-1040}
\address[3]{Department of Electrical Enginnering and Physics, Wilkes University, 84 W. South Street, Wilkes-Barre, PA 18766}

\begin{abstract}
  We study the covariance in the angular power spectrum estimates of CMB fluctuations when the primordial fluctuations are non-Gaussian. The non-Gaussian covariance  comes from a nonzero connected four-point correlation function --- or the trispectrum in Fourier space --- and can be large when long-wavelength (super-CMB) modes are strongly coupled to short-wavelength modes. The effect of such non-Gaussian covariance can be modeled through additional freedom in the theoretical CMB angular power spectrum and can lead to different inferred values of the standard cosmological parameters relative to those in $\Lambda$CDM. Taking the collapsed limit of the primordial trispectrum in the  quasi-single field inflation model as an example, we study how the six standard $\Lambda$CDM parameters shift when two additional parameters describing the trispectrum are allowed. The reduced statistical significance of the Hubble tension in the extended model allows us to combine the {\it Planck} temperature data and the type Ia supernovae data from Panstarrs with the distance-ladder measurement of the Hubble constant. This combination of data shows strong evidence for a primordial trispectrum-induced non-Gaussian covariance, with a likelihood improvement of $\Delta \chi^2 \approx -15$ (with two additional parameters) relative to $\Lambda$CDM.
\end{abstract}


\end{frontmatter}

\section{Introduction}
The statistical distribution of primordial fluctuations is a key ingredient that underpins all cosmological results obtained from analyzing the distribution of hot and cold spots of the cosmic microwave background (CMB) anisotropies,  the galaxy distribution, and lensing signal in the large-scale structure. The standard assumption predicted by simplest single-field, slow roll models of inflation, employed as a default in these analyses and affirmed by data thus far \cite{Ade:2015ava}, is that cosmic fluctuations are Gaussian random on large scales. Nevertheless, it is entirely possible that Gaussianity is violated even at large scales, and this is the subject of much interest \cite{Alvarez:2014vva}.

Searches for non-Gaussianity in the CMB have mainly focused on constraining the amplitudes of higher-order $n$-point correlation functions (generally bispectrum and trispectrum, $n=3$ and $4$) \cite{Ade:2013ydc, Ade:2015ava}. However, the presence of non-Gaussianity can also affect the two-point correlation function analyses. In particular, it is well known that the presence of a trispectrum generates additional, non-Gaussian covariance of the angular power spectrum estimators \cite{Hu:2001fa, Smith:2015uia, Adhikari:2018osh}. It is therefore interesting to explore if primordial trispectra that can generate significant level of non-Gaussian angular power spectrum covariance affect our cosmology inferences from current and future data. It would be particularly interesting if such non-Gaussian covariance, when accounted-for, helped explain the currently much-discussed discrepancy in the derived value of the Hubble constant $H_0$ between local distance-ladder type measurements ($H_0 = 74.03 \pm 1.42 \kmsMpc$; \citep{Riess:2019cxk}) and those from Planck ($H_0 = 67.4 \pm 0.5 \kmsMpc$; \citep{Aghanim:2018eyx}), at approximately $4.4\sigma$. These $H_0$ measurements and CMB likelihood were released after our analysis, for which we use Planck 2015 {temperature} likelihood \cite{Aghanim:2015xee} and a previous distance ladder $H_0$ measurement {($H_0 = 73.52 \pm 1.62 \kmsMpc$;}  \cite{Riess:2018byc}), which disagree at $3.6\sigma$.

In what follows, we adopt the quasi-single field inflationary model \cite{Chen:2009we} which features two weakly coupled scalar fields, a massless inflaton and a massive isocurvaton. In this model, primordial fluctuations have a four-point function that is large in the collapsed limit, meaning that there is coupling between small- and large-scale modes. The collapsed-limit trispectrum of the quasi-single field inflationary models has the form \cite{Assassi:2012zq}:
\begin{align}
\Bl \Phi_{\kv_1}\Phi_{\kv_2}\Phi_{\kv_3}\Phi_{\kv_4}\Br_c &= \tpc \delta(\kv_1+\kv_2+\kv_3+\kv_4) T(\kv_1, \kv_2, \kv_3, \kv_4) \nonumber \\
 T(\kv_1, \kv_2, \kv_3, \kv_4) &= 4 \taunl(\epsilon) \left(\frac{K}{\sqrt{k_1 k_3}}\right)^{-2\epsilon} P_\Phi(k_1) P_\Phi(k_3) P_\Phi(K)   \label{eq:trispectrum}
\end{align}
where $k_i$ are the four wavenumbers that define the trispectrum rectangle in momentum space, $K=|\kv_1-\kv_2|=|\kv_3-\kv_4|$, $P_\Phi(k) = (2\pi^2 A_\Phi/k^3) (k/k_0)^{n_s-1}$ is the power spectrum of potential fluctuations, and $A_\Phi$ is its amplitude. Here $\epsilon \equiv \nu-3/2<0$ is a free parameter that depends on the mass of an additional scalar field since $\nu\equiv (9/4-m_\sigma^2/H^2)^{1/2}$ where $m_\sigma$ is mass of the isocurvaton field and $H$ is the Hubble rate during inflation, while $\taunl$ is the amplitude of the collapsed four-point function. The three-point function amplitude $\fnl(\epsilon)$ in the quasi-single field model has been constrained by Planck \cite{Ade:2013ydc} (see their Fig. 26), but the corresponding constraint for the four-point amplitude $\taunl(\epsilon)$ does not currently exist. The existing bounds on $\taunl$ \cite{Ade:2013ydc, Feng:2015pva} correspond to the $\epsilon =0$ limit in Eq.~(\ref{eq:trispectrum}). In the quasi-single field model, the four-point amplitude is boosted with respect to $\fnl$: $\taunl \sim \fnl^2 /(\rho/H)^2$ for a small coupling constant $\rho$ ($\rho\ll H$) \cite{Chen:2009zp, Baumann:2011nk, Assassi:2012zq}, and therefore can be much larger than $\fnl^2$. A detection of a boosted collapsed four-point function \cite{Suyama:2007bg} would indicate the role of more than one source in generating the curvature perturbations. The effect of such four-point functions on the large-scale structure clustering has been extensively studied \cite{Tseliakhovich:2010kf, Smith:2010gx, Baumann:2012bc, Ferraro:2014jba, Adhikari:2014xua, An:2017rwo}.

\section{Effect on CMB covariance}
The higher multipoles ($\ell\gtrsim 30$) of the CMB angular power spectrum can be estimated from the maps using the pseudo-$C_\ell$ estimator,
$\hat{C}_\ell =  \sum_m a_{\ell m}^* a_{\ell m}/(2\ell+1)$, where $a_{\ell m}$ are the harmonic decomposition coefficients of the CMB map.
The full-sky covariance of this angular power spectrum is 
\begin{align}
    \cov(\hat{C}_\ell, \hat{C}_{\ell'}) = \frac{2 C_\ell^2}{2\ell+1} \delta_{\ell \ell'}  + \frac{\sum_{m,m'} \Bl a_{\ell m}^* a_{\ell m} a_{\ell' m'}^* a_{\ell' m'} \Br_c}{(2\ell+1)(2\ell'+1)}, 
    \label{eq:cov}
\end{align}
where the second term on the right-hand side, which we refer to as $\cov_{\rm NG}(\hat{C}_\ell, \hat{C}_{\ell'})$, is due to the connected part of the CMB trispectrum and is zero for Gaussian $a_{\ell m}$s.
Here we will focus on the non-Gaussian covariance contribution from a primordial trispectrum of the form Eq.~(\ref{eq:trispectrum}). The methods to compute CMB four-point functions can be found in \cite{Hu:2001fa, Okamoto:2002ik, Regan:2010cn}. If we rewrite the expression in Eq.~(\ref{eq:trispectrum}) with shifted spectral index for the power spectra ($n_s+\epsilon$ for terms with $k_i$, and $n_s-2\epsilon$ for terms with $K$), the expression and therefore the calculation of the CMB temperature angular trispectrum match those of the exact local model. The full-sky expression for the non-Gaussian covariance from the collapsed-limit of the quasi-single field trispectrum then simplifies to give:
\begin{align}
    \cov_{\rm NG}(\hat{C}_\ell, \hat{C}_{\ell'}) = \frac{9}{\pi} \taunl(\epsilon) C_{L=0}^{\rm SW}(n_s- 2\epsilon) C_{\ell}(n_s+\epsilon) C_{\ell'}(n_s+\epsilon)
    \label{eq:covng}
\end{align}
where the ${C}_\ell(n_s+\epsilon)$s are the angular power spectra evaluated at shifted values of the spectral index with all other cosmological parameters fixed. The $C_\ell$ in the expression above are the lensed harmonics \cite{Pearson:2012ba}.  The expression for the Sachs-Wolfe angular power ${C}_L^{\rm SW}$ is
\begin{align}
    {C}_L^{\rm SW}(n_s-2\epsilon) &= \frac{4\pi A_\Phi}{9} \int \frac{dK}{K} j_L^2(K r_*) \left(\frac{K}{k_0}\right)^{n_s-2\epsilon-1}  \nonumber \\ &= \frac{4\pi A_s}{25(k_0 r_*)^a} \frac{\sqrt{\pi} \Gamma(1-\frac{a}{2}) \Gamma(L+\frac{a}{2})}{4 \Gamma(\frac{3}{2}-\frac{a}{2}) \Gamma(2+L-\frac{a}{2})},
\end{align}
where  $a = n_s-2\epsilon-1, 0<a<2$,  $r_*$ is the comoving distance to the last scattering surface, and where in the second line above we have used $A_\Phi = (9/25)A_s$.

Instead of implementing the non-Gaussian covariance in data analysis, one can equivalently consider how the estimated power spectrum in a realization appears biased when the non-Gaussian covariance is not included; this can be modeled by using an additional random variable $A_0$ as follows:
\begin{align}
    \hat{C}_\ell^{\rm sky} = \hat{C}_\ell - A_0 C_{\ell}(n_s+\epsilon)
    \label{eq:Clest_correction}
\end{align}
where $\hat{C}_\ell$ is the angular power spectrum estimate for a realization with $A_0=0$, and the term $A_0 C_\ell(n_s+\epsilon)$ quantifies the bias in realizations with non-zero $A_0$. This is the ``super-sample signal" approach \cite{Li:2014jra}, previously utilized for the non-Gaussian covariance due to CMB lensing \cite{Manzotti:2014wca, Henning:2017nuy, Motloch:2018pyt}. One can explicitly check that the covariance of the $\hat{C}_\ell^{\rm sky}$ --- the right-hand side of Eq.~(\ref{eq:Clest_correction}) --- leads precisely to the desired non-Gaussian covariance in Eq.~(\ref{eq:covng}) provided the variable $A_0$ has a global distribution with mean zero and variance:
\begin{align}
    \Bl A_0^2 \Br &= \frac{9}{\pi} \taunl(\epsilon) {C}_{L=0}^{\rm SW}(n_s-2\epsilon).\label{eq:var_A}
\end{align}

\begin{figure}
    \centering
    \includegraphics[width=0.45\textwidth]{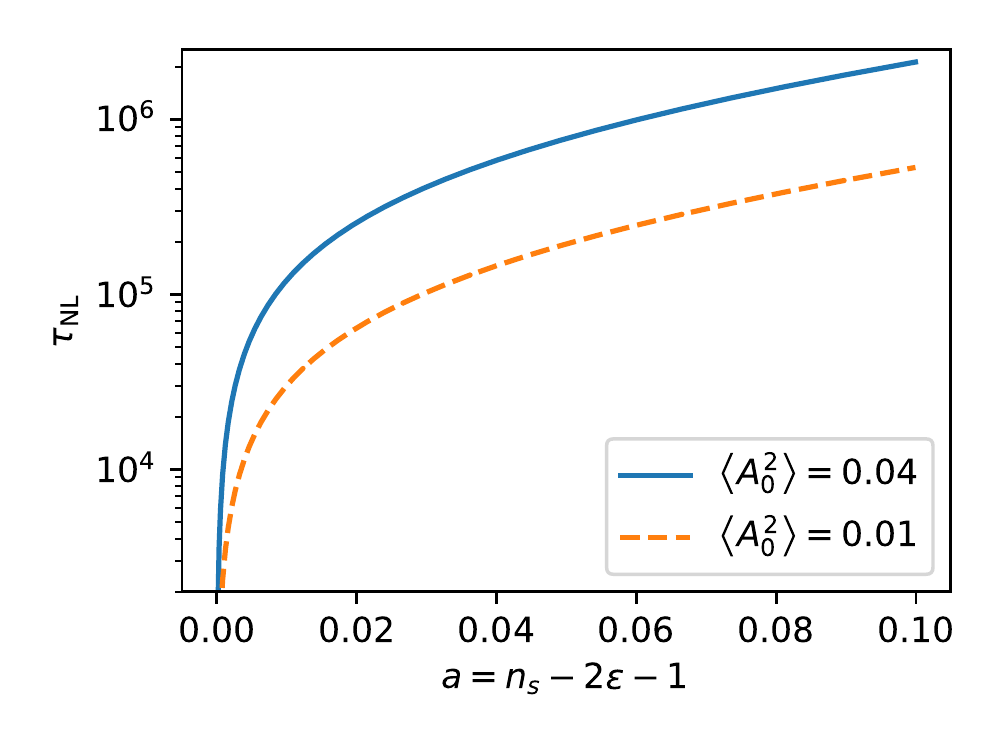}
    \caption{The trispectrum amplitude $\tau_{\rm NL}$ as a function of $a=n_s-2\epsilon-1$ resulting in the given variance of $A_0$, plotted for $\langle A_0^2 \rangle = 0.04$ and $0.01$. The expression for the variance Eq.~(\ref{eq:var_A}) diverges for  $a \leq 0$ or $n_s-2\epsilon \leq 1$. The existing constraint on $\tau_{\rm NL}$ ($\lesssim \mathcal{O}(10^4)$ \cite{Ade:2013ydc, Feng:2015pva}) implicitly assumes $\epsilon=0$ such that $a\leq 0$ for $n_s \leq 1$. As shown in the plot, for small values of $a$, $\tau_{\rm NL} \lesssim 10^4$ can produce a variance $\langle A_0^2 \rangle \simeq 0.04$ large enough to be consistent with the preferred value of $A_0\simeq -0.2$ in our data analysis.}
    \label{fig:taunl_a}
\end{figure}

The variance $\langle A_0^2 \rangle$ depends strongly on both $\taunl$ and $n_s-2\epsilon$; see Figure \ref{fig:taunl_a}. When $\epsilon=0$, the trispectrum Eq.~(\ref{eq:trispectrum}) reduces to the well-studied (local) $\taunl-$trispectrum from multifield inflation models such as the curvaton model \cite{Byrnes:2006vq, Komatsu:2010hc}. These $\epsilon=0$ models were constrained by older Planck data \cite{Ade:2013ydc, Feng:2015pva}, giving $\taunl \lesssim 10^4$. {From Figure \ref{fig:taunl_a}, we see that $\taunl \lesssim 10^4$ can produce a variance $\langle A_0^2 \rangle \simeq 0.04$ large enough to be consistent with the preferred values of $A_0$ in our analysis.} However, a direct translation of this $\taunl$ constraint to a limit on $\langle A_0^2 \rangle $ cannot be performed because of the infrared divergent term $C_{L=0}^{\rm SW}$ in Eq. (\ref{eq:var_A}) which is due to the nature of mode coupling in the local $\taunl$ trispectrum, in which arbitrarily long wavelength modes are coupled to modes that contribute to the CMB $C_\ell$s. For $\epsilon \neq 0$, no direct constraint on the trispectrum amplitude $\taunl$ exists. When such constraints are obtained through direct trispectrum measurements in the future, we can use them to further tighten the viable parameter space of $A_0, \epsilon$.

Because the CMB likelihood features the difference between data and theory, $(C_\ell-\hat{C}_\ell^{\rm sky})$, we can implement the effect in Eq.~(\ref{eq:Clest_correction}) by correcting the theoretical angular power spectrum as 
\begin{align}
    C_\ell \rightarrow C_\ell + A_0 C_{\ell}(n_s + \epsilon).
    \label{eq:Clobs}
\end{align}
 
Note that the expression for the variance in Eq.~(\ref{eq:var_A}) is altered in the presence of a cut sky, but this does not affect our data analysis for which we use Eq.~(\ref{eq:Clobs}) with $A_0$ as a free parameter. From Eq.~(\ref{eq:Clobs}) one expects a  large degeneracy between the primordial amplitude $A_s$ and $A_0$, and similarly between $n_s$ and $\epsilon$. For example, if the $C_\ell$s were linear in $A_s$, then $A_0$ and $A_s$ would be exactly degenerate. Fortunately,  power spectrum amplitude $A_s$ also controls the amount of lensing, smoothing the acoustic peaks, and thus making it possible to break the degeneracy with $A_0$ and constrain the latter parameter using CMB power spectrum.

\section{Data Analysis}
Our goal is to constrain the two parameters $A_0$ and $\epsilon$ defined in Eq.~(\ref{eq:Clobs}). We sample those parameters with priors $A_0\in [-0.5, 0.5]$ and $\epsilon \in [-0.5,0]$. We impose a hard prior cut-off at $\epsilon=0$ because it corresponds to the limit in the quasi-single field model when the isocurvaton is massless. We consider six additional, standard $\lcdm$ parameters with the following priors: physical cold dark matter density $\Omega_c h^2\in [0.001, 0.99]$; physical baryon density $\Omega_b h^2 \in [0.005, 0.1]$; power spectrum amplitude $\ln(10^{10}A_s)\in[2, 4]$; spectral index $n_s\in [0.7, 1.3]$; optical depth $\tau\in [0.01, 0.8]$; baryon peak location $100\theta_{\rm MC}\in [0.5, 10]$. We compare results for two cosmological model spaces:
\begin{itemize}
    \item $\lcdm:\{\Omega_c h^2, \Omega_b h^2, \ln(10^{10}A_s), n_s, \tau, 100\theta_{\rm MC}\}$
    \item $\slcdm: \lcdm + \{A_0, \epsilon\}$
\end{itemize}
We adopt the {\it Planck} 2015 likelihoods: high-$\ell$ \texttt{plikHM}\footnote{\texttt{plik\_dx11dr2\_HM\_v18\_TT.clik}} and low-$\ell$ \texttt{commander}\footnote{\texttt{commander\_rc2\_v1.1\_l2\_29\_B.clik}} likelihoods. We also use a Gaussian prior on the reionization optical depth $\tau$ from 2018 {\it Planck} polarization measurements (\texttt{EE+lowE}), $\tau = 0.0527\pm0.009$ \cite{Aghanim:2018eyx}. Since this $\tau$ constraint is dominated by the low-$\ell$ EE multipoles, we assume that the $\tau$ constraint is not significantly affected by the presence of modulation from a trispectrum. We use \texttt{CosmoMC} \cite{Lewis:2002ah, Lewis:2013hha} for posterior sampling and for obtaining the best-fit parameters, and the angular power spectra $C_\ell$ are computed using \texttt{camb} \cite{Lewis:1999bs, Howlett:2012mh}. 

For model comparison, we estimate the Bayes factor using the Savage-Dickey density ratio (SDDR) \cite{Trotta:2005ar}. The SDDR is an approximation to the Bayes factor for nested models. Here, the $\slcdm$ model ($M_1$) reduces to $\lcdm$ model ($M_0$) when $A_0=0$. The Bayes factor (SDDR), then, is the ratio of the posterior probability density, $p$, to the prior probability density, $\pi$, at $A_0=0$ in the $\slcdm$ model:
\begin{align}
    B_{01} = \left.\frac{p(A_0|d, M_1)}{\pi(A_0|M_1)} \right\vert_{A_0=0} \qquad ({\rm SDDR}).
\end{align}

We will  evaluate the logarithm of the Bayes factor, $\ln B_{01}$, on the modified Jeffreys' scale for the strength of evidence \cite{Trotta:2008qt}: $|\lnB|=1, 2.5, 5$ as weak, moderate and strong evidence in favor of $M_1$ respectively.

\begin{table}
\centering
\caption{Best-fit values of the parameters and the marginalized 1D 95\% limits. We also list the improvement in $\chi^2$ with respect to the $\lcdm$ model: $\Delta \chi^2 = \chi^2_{\rm best\-fit}(\slcdm)-\chi^2_{\rm best\-fit}(\lcdm)$, and an approximation to the Bayes factor, $\lnB$, calculated using the Savage-Dickey density ratio.}
\label{table:bestfit}
\small
\begin{tabular*}{0.48\textwidth}{l@{\extracolsep{\fill}}  c c | c c}
\hline \hline
           & \multicolumn{2}{c|}{TT+$\tau-$prior} & \multicolumn{2}{c}{TT+$\tau-$prior+$H_0$+SNIa} \\
 Parameter &  Best fit &  95\% limits & Best fit &  95\% limits \\
\hline
{\boldmath$\Omega_b h^2   $} & $0.02269                   $ & $0.02256^{+0.00062}_{-0.00059}$ & $0.02295                   $ & $0.02286^{+0.00050}_{-0.00050}$ \\

{\boldmath$\Omega_c h^2   $} & $0.1169                    $ & $0.1179^{+0.0050}_{-0.0051}$ & $0.11430                   $ & $0.1148^{+0.0035}_{-0.0035}$ \\

{\boldmath$100\theta_{MC} $} & $1.04136                   $ & $1.0412^{+0.0011}_{-0.0010}$ & $1.04177                   $ & $1.04166^{+0.00093}_{-0.00091}$ \\

{\boldmath$\tau           $} & $0.0534                    $ & $0.053^{+0.018}_{-0.018}   $ & $0.0527                    $ & $0.054^{+0.018}_{-0.018} $\\

{\boldmath$A_0       $}& $-0.190                    $ & $-0.15^{+0.14}_{-0.13}$  & $-0.238                    $ & $-0.21^{+0.12}_{-0.10}     $\\

{\boldmath$\epsilon$} & $-0.095                    $ & $> -0.320                  $ & $-0.058                    $ & $> -0.200                  $\\

{\boldmath${\rm{ln}}(10^{10} A_s)$}& $3.246                     $ & $3.20^{+0.15}_{-0.14}      $ & $3.301                     $ & $3.27^{+0.12}_{-0.13}      $\\

{\boldmath$n_s            $} & $0.9515                    $ & $0.950^{+0.025}_{-0.028}   $ & $0.9639                    $ & $0.954^{+0.028}_{-0.030}   $\\
{\boldmath$H_0$} & $68.89                     $ & $68.4^{+2.5}_{-2.3} $ & $70.18                     $ & $69.9^{+1.7}_{-1.7}        $ \\
\hline
$\Delta \chi^2$    &     -7.8 & & -15.0 \\
{$\lnB$} & -0.6 & & -3.8 \\ 
\hline
\end{tabular*}
\end{table}

\begin{figure*}
    \centering
    \includegraphics[width=0.85\textwidth]{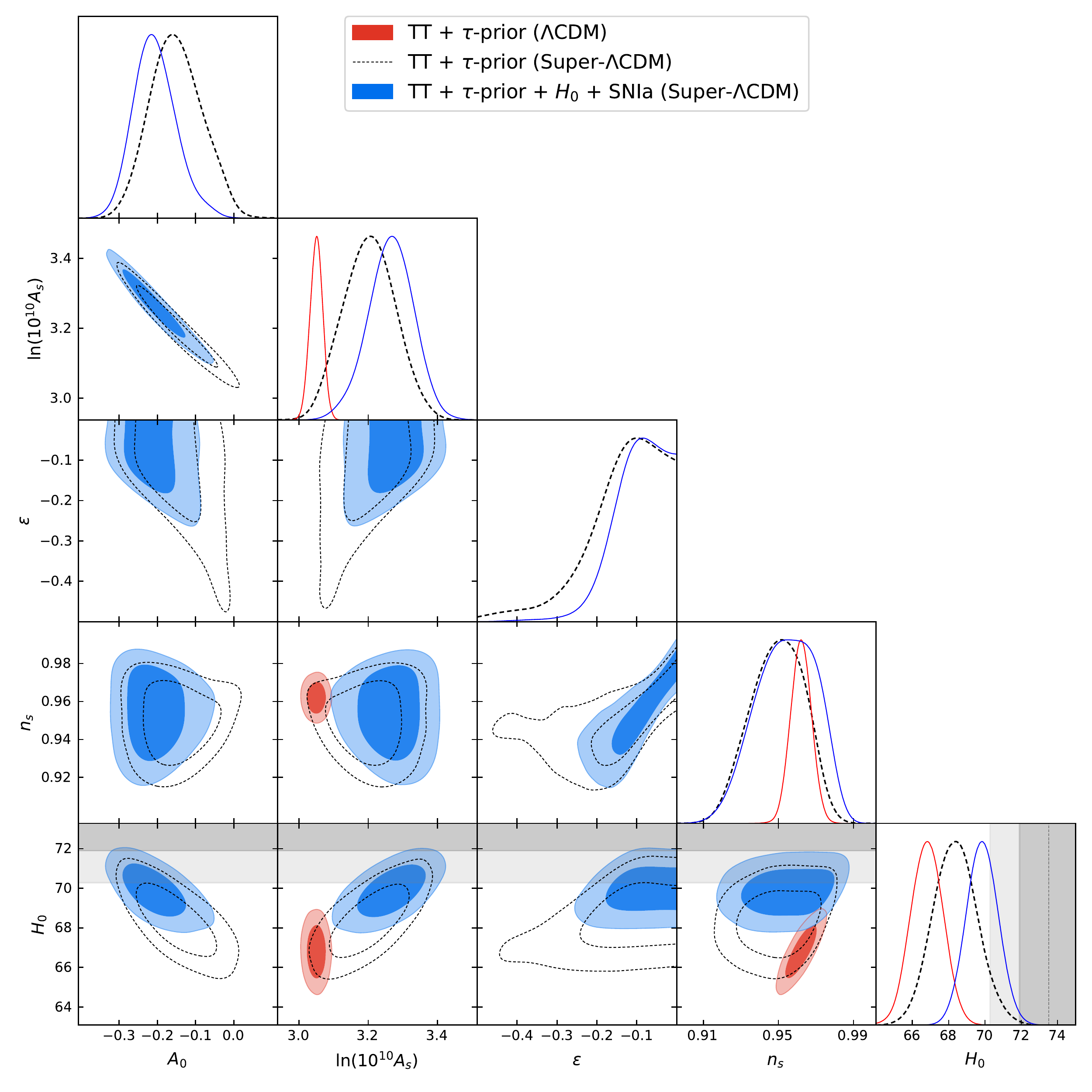}
    \caption{Marginalized 1D and 2D posterior distributions for the parameters describing the primordial fluctuations $\left\{\ln(10^{10}A_s), n_s, A_0, \epsilon \right\}$, along with $H_0$, for different choices of data and models. The red contours show the results for the base $\Lambda$CDM model using Planck data. Allowing for the non-Gaussian covariance significantly broadens and shifts the constraints on the primordial amplitude and spectral index (thin-line contours). Adding $H_0$ and SNIa data helps in constraining the parameters in the Super-$\Lambda$CDM model (blue contours). {The gray line and bands in the lower panels show the measurement and uncertainty (1 and 2 $\sigma$) of the distance-ladder Hubble constant measurement from \cite{Riess:2018byc}.}}
    \label{fig:triangle1}
\end{figure*}

\section{Results}

\begin{figure*}
    \centering
    \includegraphics[width=0.7\textwidth]{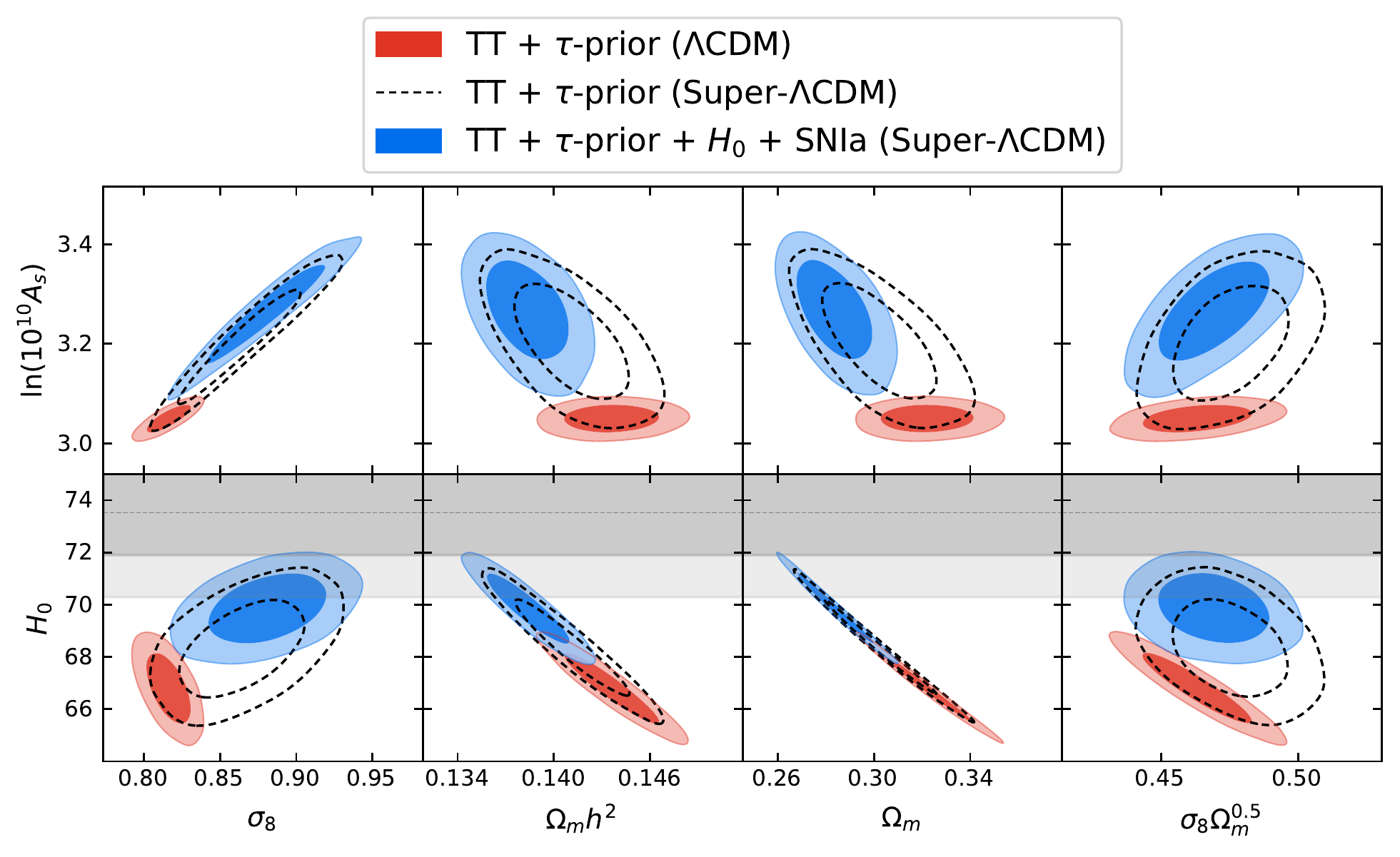}
    \caption{Posterior distributions for other parameters of interest, $\Omega_m$, $\Omega_m h^2$, $\sigma_8$ and $\sigma_8 \Omega_m^{0.5}$; see also discussion in the text.}
    \label{fig:rectangle1}
\end{figure*}

While a primordial trispectrum affects the observations which depend on density fluctuations (CMB power spectrum, galaxy two-point functions, cluster counts etc), observations which directly probe cosmic expansion are unaffected. Therefore, we can add the Type Ia supernovae (SNIa) data from the Pantheon sample \cite{Scolnic:2017caz} and the distance-ladder measurement of the Hubble constant \cite{Riess:2018byc}, without making any changes to the theoretical predictions or the likelihoods. Our primary data set, therefore, is: \texttt{TT + $\tau-$prior + $H_0$ + SNIa}. The 1D and 2D marginalized posterior distributions of cosmological parameters for $\lcdm$ and $\slcdm$ models are shown in Figure \ref{fig:triangle1} {and Figure \ref{fig:rectangle1}}; the corresponding best-fit parameter values and the 1D marginalized $95\%$ limits are given in Table \ref{table:bestfit}. Next, we discuss the results for different data combinations.

\subsection{Planck-only (TT+$\tau-$prior) data}
The $\slcdm$ fit to the {\it Planck}-TT + $\tau-$prior data (thin-line contours in Figs.~\ref{fig:triangle1}) prefers about $20\%$ larger  amplitude of fluctuations at large scales relative to $\lcdm$ (red contours), that is, $A_0\simeq -0.2$ and the $A_s$ correspondingly higher by about 20\%. The improvement in fit relative to $\lcdm$ is $\Delta \chi^2 = -7.8$. The Bayes factor, $|\lnB|=0.6 <1$, provides ``inconclusive" evidence in favor of $\slcdm$ for this data combination according to the modified Jeffreys' scale \cite{Trotta:2008qt}. The preference for more power at large scales is accompanied by increased lensing of the CMB power spectrum, which has been observed in Planck TT data via the preference for $A_{\rm lens}>1$, where $A_{\rm lens}$ is a phenomenological lensing amplitude parameter. The $\slcdm$ model, therefore, provides a physical explanation for the preference for increased lensing in {\it Planck} TT data. The corresponding $H_0$ constraint is both broader and shifts towards larger values thereby reducing the tension with local measurements of $H_0$. The discrepancy in $H_0$ (1D marginalized) reduces to $2.5\sigma$ in $\slcdm$ compared to the $3.6\sigma$ discrepancy in $\lcdm$. We can, therefore, combine {\it Planck} and the local $H_0$ data to further constrain the $(A_0,\epsilon)$ parameters of the $\slcdm$ model.

\subsection{TT+$\tau$-prior + $H_0$ + SNIa data}
This is the fiducial data combination employed in this paper. As we can see in the posterior distributions plotted in Figure \ref{fig:triangle1}, combining the distance-ladder $H_0$ and SNIa Pantheon data with Planck (blue contours) helps improve the constraints on $A_0$ and $\epsilon$, and leads to the improvement in the fit of $\Delta\chi^2=-15$ relative to the equivalent $\lcdm$ case, thus favoring the extended model with two additional parameters by about $3.5\sigma$. The Bayes factor $|\lnB|=3.8$ suggests a ``moderate-to-strong" preference for $\slcdm$. The 1D marginalized constraint on the modulation amplitude is $A_0=-0.21^{+0.12}_{-0.11}$ ($95 \%$ limits) and the corresponding constraint on the Hubble constant is $H_0 = 69.9 \pm 1.7$. 

$\slcdm$ model allows additional freedom in the overall amplitude of the CMB power spectrum. As shown in Figure \ref{fig:triangle1}, this results in a large upward shift in the amplitude of primordial fluctuations $A_s$ (and thus in $\sigma_8$). The consequent positive correlation between $A_s$ and $H_0$ in the $\slcdm$ model can be understood in terms of well-known CMB degeneracies as follows. First, $\Omega_m h^2$ is anti-correlated with $A_s$ as large values of $\Omega_m h^2$ lower the overall amplitude of the CMB power spectrum. Second, tight constraints on the peak locations and relative heights pin down the parameter combination $\Omega_m h^3$ \cite{Percival:2002gq}, resulting in an anti-correlation between $\Omega_m h^2$ and $H_0$. It then follows that data prefer a larger $H_0$ when a higher amplitude of primordial fluctuations $A_s$ than in $\lcdm$ is favored, which in turn happens when $A_0<0$.

The near-$20\%$ increase in $A_s$ and $\sigma_8$ does not exacerbate the tension between the amplitude of matter fluctuations are measured by the CMB and weak lensing surveys. Weak lensing surveys are sensitive to a specific combination of matter density $\Omega_m$ and amplitude of mass fluctuations $\sigma_8$, $S_8 \equiv \sigma_8 (\Omega_m/0.3)^{0.5}$ (e.g.\ \cite{Abbott:2017wau}). As we  see in Figure \ref{fig:rectangle1}, the increase in $\sigma_8$ does not appreciably change the constraint on the parameter combination $S_8$ in $\slcdm$. It is important to note, however, that the weak lensing data might be sensitive to the changes in covariance matrix in $\slcdm$. Therefore careful further analysis is required to compare $S_8$ measured by weak lensing and CMB in the $\slcdm$ model.

\subsection{TT+$\tau-$prior+$H_0$+SNIa+BAO data}
A primordial trispectrum with signal in the collapsed limit has two major effects on the galaxy clustering data: (i) scale-dependent bias at large scales in the galaxy-galaxy power spectrum \cite{Dalal:2007cu, Baumann:2012bc}, and (ii) super-sample effect similar to the CMB case which can bias the power spectrum estimate of a given survey. The scale-dependent bias mostly affects the broadband shape of the power spectrum, and should be largely --- though not necessarily completely --- removed in BAO analyses which remove systematics by subtracting smooth polynomials from the $P(k)$ wiggles \cite{Anderson:2013zyy}. The super-sample effect {will also} change the amplitude and scale dependence of the template used to fit the BAO feature.

Because accurate modeling of the shift of BAO peaks in the presence of primordial non-Gaussianity is a complex task, simply adopting the reported BAO distance measurements can not be considered as wholly reliable. Nevertheless we carried out two preliminary tests. First, we found that the BAO peak-shifts in the presence of $\taunl$ values favored by the data are generally much smaller than the corresponding BAO peak-location statistical errors. Second, we also ran an exploratory analysis where we simply combined the reported BAO distance measurements \cite{Ross:2014qpa,Gil-Marin:2015nqa,Beutler:2011hx} with the full dataset from our main analysis; in that case the $\slcdm$ model is favored over the $\lcdm$ by $\Delta \chi^2 \simeq -13$. These two findings lead us to believe that the complete BAO analysis, when  fully calibrated for non-Gaussian models, will not appreciably change the results from TT + $\tau-$prior + $H_0$ + SNIa presented above.

\section{Summary and Conclusion}
In this work, we use the angular power spectrum measurements of CMB temperature fluctuations from the {\it Planck} satellite in combination with the reionization optical depth estimate from CMB polarization, the distance-ladder measurement of the Hubble constant, and the Pantheon supernova sample to constrain the cosmological parameters in the presence of ``super-sample" fluctuations predicted by some classes of inflationary models, which we call the $\slcdm$ model. These models require a nonzero primordial trispectrum, which generates a non-Gaussian covariance of the angular power spectrum. We employ the super-sample signal technique to consider the effect of the non-Gaussian covariance on the measured CMB power spectrum. We find that the lensing of the CMB is instrumental in breaking parameter degeneracies to constrain the amplitude and spectral slope of the super-sample effect. The $3.6\sigma$ tension in the Hubble constant decreases to $2.5\sigma$ when the inflationary super-sample effect is included so that we can combine the CMB data with the distance-ladder measurement of the Hubble constant in the $\slcdm$ model. In this case, the improvement in likelihood over $\lcdm$ is substantial at the robust statistical level of $\Delta \chi^2 = -15$ {(with two additional free parameters)}. The improved fit is driven by an upward shift of the Planck-inferred Hubble constant in this class of models. This {super-sample} explanation is, in our view, at least as appealing as extant new-physics explanations for the Hubble tension \cite{DiValentino:2017rcr,Poulin:2018cxd, Vattis:2019efj,Pandey:2019plg,Kreisch:2019yzn, Agrawal:2019lmo}, and is equally or more statistically favored than them. 

The primordial trispectrum responsible for the super-sample effect can be probed directly in the four-point function of Planck data \cite{Bordin:2019tyb}. Current searches for non-Gaussianity through higher-order $n$-point functions in the Planck data have generally fixed the $C_\ell$ to that of the best-fit $\lcdm$ model. Given the possibility of large deviations from the best-fit $\lcdm$ in the presence of a primordial trispectrum, it is important to perform analyses of {the CMB} three- and four-point functions in combination with the power-spectrum analysis using techniques applied to this work. Encouraging results obtained here also motivate further studies of large-scale structure observables, such as the BAO and the weak lensing power spectrum, in the presence of primordial non-Gaussianity.

\textbf{Note added:} The data analyses presented in this work were performed when the most recent Planck 2018 CMB likelihood (including robust likelihood for polarization) and the Riess et.\ al.\ (2019) Hubble constant measurements, which have increased the parameter tension in $H_0$ compared to the data we have used, were not released. Updated analyses using newer data are underway and will be presented in a future publication.

\section*{Acknowledgments}
The authors are supported by NASA under contract 14-ATP14-0005. DH is also supported by DOE under Contract No. DE-FG02-95ER40899. This research was supported in part through computational resources and services provided by Advanced Research Computing at the University of Michigan, Ann Arbor. Some of the computations in this work used the Extreme Science and Engineering Discovery Environment (XSEDE), which is supported by National Science Foundation grant number ACI-1548562. We thank Wayne Hu, Pavel Motloch, Sarah Shandera and Yi Wang for providing valuable comments on the manuscript.


\end{document}